\documentclass[aps,prc,superscriptaddress,twocolumn,amssymb,amsmath,showpacs]{revtex4}

\usepackage{graphicx}

\begin{document}

\title{$\gamma$ ray spectroscopy of $^{19}$C via single neutron knock-out 
reaction}

%\thanks{}

\author{Zs.~Vajta}
\affiliation{Institute for Nuclear Research of the 
Hungarian Academy of Sciences, P.O. Box 51, Debrecen, H-4001, Hungary}
\author{Zs.~Dombr\'adi}
\affiliation{Institute for Nuclear Research of the 
Hungarian Academy of Sciences, P.O. Box 51, Debrecen, H-4001, Hungary}
\author{Z.~Elekes}
\affiliation{Institute for Nuclear Research of the 
Hungarian Academy of Sciences, P.O. Box 51, Debrecen, H-4001, Hungary}
\author{T.~Aiba}
\affiliation{Niigata University, Niigata 950-2181, Japan}
\author{N.~Aoi}
\affiliation{The Institute of Physical and 
Chemical Research (RIKEN), 2-1 Hirosawa, Wako, Saitama 351-0198, Japan}
\author{H.~Baba}
\affiliation{The Institute of Physical and 
Chemical Research (RIKEN), 2-1 Hirosawa, Wako, Saitama 351-0198, Japan}
\author{D.~Bemmerer}
\affiliation{Helmholtz-Zentrum Dresden-Rossendorf, Bautzner Landstrasse 400, 
01328 Dresden (Rossendorf), Germany}
\author{Zs.~F\"ul\"op}
\affiliation{Institute for Nuclear Research of the 
Hungarian Academy of Sciences, P.O. Box 51, Debrecen, H-4001, Hungary}
\author{N.~Iwasa}
\affiliation{Tohoku University, Sendai, Miyagi 9808578, Japan}
\author{\'A.~Kiss}
\affiliation{E\"otv\"os Lor\'and University, 1117 Budapest,
P\'azm\'any P\'eter s\'et\'any 1/A, Hungary}
\author{T.~Kobayashi}
\affiliation{Tohoku University, Sendai, Miyagi 9808578, Japan}
\author{Y.~Kondo}
\affiliation{Tokyo Institute of Technology,
2-12-1 Oh-Okayama, Megro, Tokyo 152-8551, Japan}
\author{T.~Motobayashi}
\affiliation{The Institute of Physical and 
Chemical Research (RIKEN), 2-1 Hirosawa, Wako, Saitama 351-0198, Japan}
\author{T.~Nakabayashi}
\affiliation{Tokyo Institute of Technology,
2-12-1 Oh-Okayama, Megro, Tokyo 152-8551, Japan}
\author{T.~Nannichi}
\affiliation{Tokyo Institute of Technology,
2-12-1 Oh-Okayama, Megro, Tokyo 152-8551, Japan}
\author{H.~Sakurai}
\affiliation{The Institute of Physical and 
Chemical Research (RIKEN), 2-1 Hirosawa, Wako, Saitama 351-0198, Japan}
\author{D.~Sohler}
\affiliation{Institute for Nuclear Research of the 
Hungarian Academy of Sciences, P.O. Box 51, Debrecen, H-4001, Hungary}
\author{S.~Takeuchi}
\affiliation{The Institute of Physical and 
Chemical Research (RIKEN), 2-1 Hirosawa, Wako, Saitama 351-0198, Japan}
\author{K.~Tanaka}
\affiliation{The Institute of Physical and 
Chemical Research (RIKEN), 2-1 Hirosawa, Wako, Saitama 351-0198, Japan}
\author{Y.~Togano}
\affiliation{The Institute of Physical and 
Chemical Research (RIKEN), 2-1 Hirosawa, Wako, Saitama 351-0198, Japan}
\author{K.~Yamada}
\affiliation{The Institute of Physical and 
Chemical Research (RIKEN), 2-1 Hirosawa, Wako, Saitama 351-0198, Japan}
\author{M.~Yamaguchi}
\affiliation{The Institute of Physical and 
Chemical Research (RIKEN), 2-1 Hirosawa, Wako, Saitama 351-0198, Japan}
\author{K.~Yoneda}
\affiliation{The Institute of Physical and 
Chemical Research (RIKEN), 2-1 Hirosawa, Wako, Saitama 351-0198, Japan}

\date{\today}

\begin{abstract}
The one neutron knock-out reaction $^{1}$H($^{20}$C,$^{19}$C$\gamma$) was studied at RIKEN using the DALI2 array. A $\gamma$ ray transition was observed at 198(10)~keV. Based on the comparison between the experimental production cross section and theoretical predictions, the transition was assigned to the decay of the $3/2_1^+$ state to the ground state.
\end{abstract}

\pacs{23.20.Js, 25.60.-t, 27.30.+t, 29.30.Kv}
%\keywords{}

\maketitle

\section{Introduction}

Construction of radioactive ion beam facilities opened new ways in nuclear structure studies. Neutron-rich nuclei far from the valley of stability became experimentally reachable in the past two decades. The neutron-rich carbon isotopes showing interesting phenomena like one~\cite{nakamura} and two neutron halo~\cite{tanaka_22c,kobayashi_22c}, neutron decoupling~\cite{elekes_16c,elekes_20c}, weakening of the neutron-neutron effective interaction~\cite{msu40si}, development of the $N$~=~16 subshell closure and disappearance of the $N$~=~14 one~\cite{stanoiu3} were in the focus for a long time.

Since its identification by Bowman et al.~\cite{bowman}, probably $^{19}$C was the most investigated nucleus in the lower mass region of isotopes. It attracted attention as a candidate of a one-neutron halo nucleus due to its low binding energy and spin 1/2$^+$ ground state suggested by shell model calculations. The large interaction~\cite{ozawa0} and Coulomb dissociation~\cite{nakamura} cross sections supported this assumption. The momentum distribution probed in different ways by several groups~\cite{bazin,maddalena,baumann,kanungo,chiba,marques} were consistent with the halo nature and the ground state spin 1/2$^+$ assignment, but 3/2$^+$ and 5/2$^+$ spins were not completely excluded as discussed in Refs.~\cite{baumann,kanungo}. Even though, there is a consensus that the dominant character of the $^{19}$C ground state is 1s$_{1/2} \otimes $0$^+$ on the basis of the observed spectroscopic factors and the absolute break up cross sections. The halo nature and the spin 1/2$^+$ ground state assignment was confirmed in a recent experiment, too~\cite{yamaguchi}.

Concerning the excited states of $^{19}$C, two $\gamma$ rays in the $^{19}$C(p,p') reaction were observed at 72(4)~keV  and 197(6)~keV energies~\cite{elekes}, which were assigned to the $5/2^+ \rightarrow 3/2^+ \rightarrow 1/2^+$ decay sequence. The existence of the higher energy transition was confirmed in a multi nucleon removal reaction~\cite{stanoiu3}, where a 201(15)~keV transition was observed. An unbound excited state was revealed at 1.46(10)~MeV in the (p,p') process via detection of the emitted neutrons~\cite{satou}. Recently, another unbound state was observed at 653(95)~keV in a multi proton removal reaction via detection of the emitted neutrons from the unbound $^{19}$C states~\cite{thoennessen}. The state at 1.46~MeV excitation energy was assigned to a 5/2$^+$ state on the basis of an angular distribution measurement~\cite{satou}. It may have an $s_{1/2} \otimes 2^+$ core excited  configuration according to shell model calculations~\cite{thoennessen}. On the other hand, the presence of three low energy excited states contradicts to the shell model expectations. To resolve this contradiction, we studied the single neutron knock out reaction from $^{20}$C since, according to shell model + Glauber model calculations for this reaction, only the $s_{1/2}$ and $d_{5/2}$ single particle states were expected to be excited with a large cross section~\cite{ozawa,kobayashi}. 

\section{Experimental details}

The experiment was performed at the Nishina Center for Accelerator-Based Science located in RIKEN, Japan~\cite{elekes_20c}. As a first step, a stable $^{40}$Ar beam of 700~pnA was produced by using the RILAC linear accelerator coupled to the RRC cyclotron. This ion beam, the energy of which was 63~MeV/nucleon, hit a target made of $^{181}$Ta with a thickness of 0.2~mm. The $^{40}$Ar particles were fragmented in the target. The reaction products were purified by the RIPS radioactive ion separator~\cite{rips}. This purification was performed on the basis of the different magnetic rigidity (B$\rho$) of the isotopes by applying two dipole magnets between which a wedged-shape aluminum degrader of 221~mg/cm$^2$ thickness was placed for inducing dispersion at the first focal plane (F1). The momentum acceptance of the fragment separator was set to the maximum 6\%. The total intensity of the radioactive ion beam was about 100~particle/s~(pps). The following main species were included in the beam $^{17}$B (11.32\%), $^{19}$C (18.02\%), $^{20}$C (9.77\%), $^{21}$N (45.76\%) and $^{22}$N (12.63\%). These were identified by their energy loss ($\Delta$E), time-of-flight (ToF) and B$\rho$~\cite{pid}. $\Delta$E was determined by a silicon detector with an area of 5~cm$\times$5~cm and a thickness of 0.1~mm located at the second focal plane (F2) while the ToF was measured between two plastic scintillators put 6~m away from each other at the F2 and F3 focal planes. The beam trajectory was also monitored on an event-by-event basis by parallel plate avalanche counters (PPAC) at F2 and F3. A complete separation of the beam constituents could be achieved.

The radioactive ion beam was transported to a secondary target of liquid hydrogen of 190~mg/cm$^2$ cooled down to 22~K~\cite{htarget}. The mean energy of the $^{20}$C particles in the middle of the target was around 50~MeV/nucleon. The isotopes created by neutron knock-out reactions were identified by their ToF, $\Delta$E and total energy (E). A plastic scintillator of 1~mm thickness was put 80~cm downstream of the target which served two purposes: it measured $\Delta$E and gave the start signal for ToF. The stop signal for ToF was provided by an array of 16 plastic scintillators of 6~cm thickness. The length of each bar was 1~m, thus a total area of about 1~m$\times$1~m was covered, which ensured a full coverage (6.5$^\circ$ in laboratory system) of the outgoing reaction products. Since the isotopes fully stopped in the scintillators, they were also used to determine E.

\begin{figure}
\includegraphics[scale=0.42,clip=]{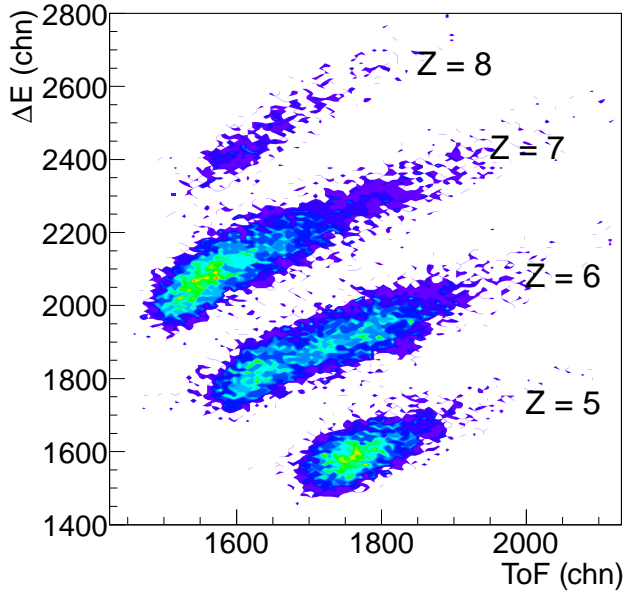}
\caption{\label{fig:ZID} (Color online) Energy loss and time-of-flight of the reaction products plotted against each other.}
\end{figure}

\begin{figure}
\includegraphics[scale=0.42,clip=]{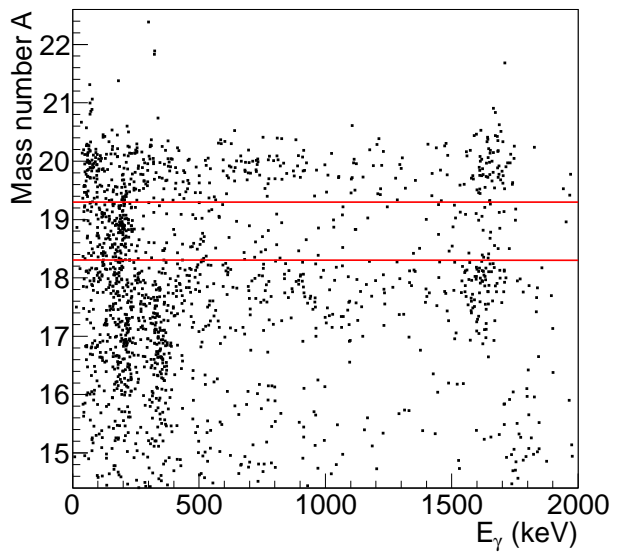}
\caption{\label{fig:EVSA} (Color online) Mass number and \textit{M}=1 Doppler-corrected $\gamma$-ray energy plotted against each other. The red lines indicate the range of projection for $^{19}$C.}
\end{figure}

The Z identification was complete and based on the combination of $\Delta$E and ToF presented in Fig.~\ref{fig:ZID}. The mass separation was performed by using the two-dimensional plot of ToF and E. The resolving power was enough for a complete distinction between mass values differing by two units, however we had some leakage between adjacent isotopes. This can be seen in Fig.~\ref{fig:EVSA}. Nevertheless, this did not imply a problem since the odd carbon isotopes have low energy $\gamma$ rays (below 600~keV) while even ones emit relatively high energy $\gamma$'s (above 1~MeV).

The deexcitation $\gamma$ rays were observed by an array of scintillators called DALI2~\cite{dali} arranged in a ball-like structure around the target. The DALI2 detector system contained 160 NaI(Tl) crystals in 16 layers, thus the setup covered a range of polar angles in the laboratory frame between 15$^\circ$ and 160$^\circ$. In order to determine the detection efficiency for $\gamma$ rays between 100~keV and 250~keV a \textsc{Geant4} simulation was constructed which provided 54~\% efficiency at 200~keV. The simulation showed good agreement with experimental data available by radioactive sources at around 1~MeV.

The energy of the $\gamma$ rays were corrected for the Doppler effect by using the known average velocity of the beam constituents in the middle of the target and the position of the NaI(Tl) detectors. The time signals for each of the member of the DALI2 setup as well as the hit multiplicity of the array (\textit{M}) were recorded. 

\section{Results and Discussion}

\begin{figure}
\includegraphics[scale=0.44,clip=]{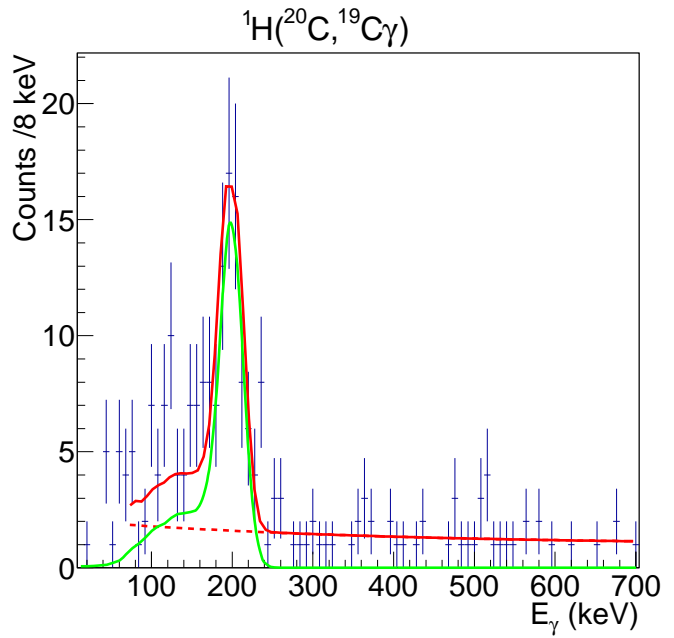}
\caption{\label{fig:spectrum} (Color online) Doppler-corrected spectra of $\gamma$ rays emerging from neutron knock-out reaction $^{1}$H($^{20}$C,$^{19}$C$\gamma$). The solid red line is the final fit including the spectrum curves from \textsc{Geant4} simulation indicated with a solid green line and an additional smooth first-degree polynomial background plotted in dashed red line.}
\end{figure}

The \textit{M}=1 $\gamma$ ray spectrum for $^{19}$C from neutron knock-out reaction obtained  by use of the prompt time gate is presented in Fig.~\ref{fig:spectrum}. A peak can be seen at 198(10)~keV. The indicated uncertainty of the peak position is the square root of the sum of the squared uncertainties including two main errors namely the statistical one and the one due to the uncertainty in Doppler correction. We took into account the uncertainties of the average velocity of the beam and detection angle of $\gamma$ rays. The obtained energy is in accordance with the ones observed in earlier studies: 197(6)~keV~\cite{elekes} and 201(15)~keV~\cite{stanoiu3}. 

The spectrum was fitted with the response of the array from a \textsc{Geant4} simulation~\cite{geant4} plus a smooth exponential background. First, the function of peak width versus $\gamma$ ray energy was derived based on known $\gamma$ rays at 217(7)~keV and 342(10)~keV of $^{17}$C and 1601(47)~keV of $^{18}$C. Using this function, a \textsc{Geant4} simulation of the \textit{M}=1 spectrum was created, the simulated response curve at 198~keV with the smooth first-degree polynomial background fitted the experimental data points well and provided the peak height. The net counts in the peak for the spectra with liquid hydrogen were obtained from the fit. 

The total cross section for the production of this $\gamma$ ray was deduced taking into the DALI2 efficiency. It was found that in the $^{1}$H($^{20}$C,$^{19}$C$\gamma$) reaction the neutron removal cross section to the first excited state is $\sigma$(198, H)~=~4.18(85)~mb, much smaller than the 24(4)~mb inelastic scattering cross section for the $^{20}$C(p,p') process~\cite{elekes_20c}, and much smaller than expected for neutron removal cross section for such a weakly bound system. According to the calculation of Ozawa et al.~\cite{ozawa} the total cross section for the one neutron removal reaction from $^{20}$C on a proton target is 127~mb, which is shared mainly between the $s_{1/2}$ and $d_{5/2}$ states produced with 52~mb and 75~mb cross sections, respectively. The production cross section for the 3/2$_1^+$ state was calculated to be 3.6~mb, which is consistent with the experimental cross section obtained in the present study for the 198~keV state. This finding confirms the spin assumptions made in Refs.~\cite{elekes,thoennessen}. On the other hand, there is no sign for a much stronger transition which should feed the 198~keV state if we had a higher energy bound 5/2$^+$ state. If we assume that the $\gamma$ ray at 72~keV exists and connects the 5/2$_1^+$ state to the 3/2$_1^+$ one than the peak at 198(10)~keV from present experiment should be roughly an order of magnitude larger than the observed peak area due to the cascade feeding. Furthermore, we should see the 72~keV line in the \textit{M}=1 $\gamma$ ray spectrum with about 5 times larger intensity than the 198 keV line in the present spectrum considering the difference of production cross section~\cite{ozawa} for 3/2$_1^+$ state (3.6~mb) and 5/2$_1^+$ one (75~mb), the efficiency of DALI2 at 72~keV (approx 20\%) and 198~keV (approx 54\%) and the fact that a significant part of the intensity would be shifted to the multiplicity 2 part of the spectrum. Even a weak second gamma ray can be ruled out up to the neutron threshold as it is seen in Fig.~\ref{fig:spectrum}. The slight increase in the spectrum around 110~keV could correspond to a peak at the level of significance of 1.22. Thus, it is regarded as a statistical fluctuations in the background. Thus, we can completely exclude the possibility of the existence of the bound 5/2$^+$ state above the 3/2$^+$ one. Having a lower energy $d_{5/2}$ state would result in an isomeric state, the existence of which was expelled by Kanungo et al.~\cite{kanungo}. The experimentally observed production cross sections of $^{19}$C in the $^{20}$C$-$n reaction observed in Refs.~\cite{ozawa,kobayashi,yamaguchi} are much lower than they would be having both single particle states bound.

By means of the spectroscopic factor from the absolute Coulomb cross section measurement~\cite{maddalena} the ground state spin and parity of $^{19}$C is well established to be 1/2$^+$. In addition, the excited 5/2$^+_1$ and 5/2$^+_2$ states at 0.653(95)~MeV and 1.46(10)~MeV were proven unbound via neutron removal reactions~\cite{thoennessen} and proton inelastic scattering~\cite{satou}. Based on the WBP shell-model calculation~\cite{warburton,dieperink} both the 5/2$^+_1$ and the the 3/2$^+_1$ excited state are predicted to be bound. The energy of the 5/2 state is expected to be even at lower energy than the 3/2 one. As it has been pointed out by Kanungo \textit{et al.}~\cite{kanungo}, the decay of the 5/2$^+_1$ state to the ground state is expected to be strongly hindered, while that of the 3/2$^+_1$ is prompt. Due to the prompt nature of our low-energy $\gamma$ transition and the Weisskopf estimate of the half lives of 
deexciting states (see Fig. 1 in Ref.~\cite{kanungo}) a low-energy prompt transition should have M1 nature. This is consistent with the above conclusion: the $\gamma$ transition, which we see, connects the first excited state of spin 3/2$^+$ and the ground state of spin 1/2$^+$. 

\section{Summary}

The one neutron knock-out reaction $^{1}$H($^{20}$C,$^{19}$C$\gamma$) was studied at RIKEN using the DALI2 array. A weak $\gamma$-ray transition was observed at 198(10)~keV. Based on the comparison between the experimental production cross section and a theoretical prediction, the transition was assigned to the decay of the $3/2_1^+$ state to the ground state. If we had a bound 5/2$^+$ state above the 3/2$^+$ one, two transitions should have been observed with 20 times larger intensity than the intensity of the present 198~keV transition. Combining this observation with the conclusion made by Kanungo~\cite{kanungo}, i.e, non-observation of isomeric states in $^{19}$C, the existence of a bound $d_{5/2}$ state in $^{19}$C can be excluded.

\begin{acknowledgments}
We would like to thank the RIKEN Ring Cyclotron staff for their assist during the experiment. The present work was partly supported by the Grant-in-Aid for Scientific Research (No. 1520417), by the Ministry of Education, Culture, Sports, Science and Technology, by OTKA contract No. K100835 and by the European Union and the State of Hungary, co-financed by the European Social Fund in the framework of T\'AMOP-4.2.4.A/ 2-11/1-2012-0001 ‘National Excellence Program’.
\end{acknowledgments}

%\bibliography{draft_19C_150330}

\begin{thebibliography}{100}

\bibitem{nakamura} T.\ Nakamura, N.\ Fukuda, T.\ Kobayashi, N.\ Aoi, H.\ Iwasaki, T.\ Kubo, A.\ Mengoni, M.\ Notani, H.\ Otsu, H.\ Sakurai, \textit{et al.}, Physical Review Letters \textbf{83}, 1112 (1999).

\bibitem{tanaka_22c} K.\ Tanaka, T.\ Yamaguchi, T.\ Suzuki, T.\ Ohtsubo, M.\ Fukuda, D.\ Nishimura, M.\ Takechi, K.\ Ogata, A.\ Ozawa, T.\ Izumikawa, \textit{et al.}, Physical Review Letters \textbf{104}, 062701 (2010).

\bibitem{kobayashi_22c} N.\ Kobayashi, T.\ Nakamura, J.\ A.\ Tostevin, Y.\ Kondo, N.\ Aoi, H.\ Baba, S.\ Deguchi, J.\ Gibelin, M.\ Ishihara, Y.\ Kawada, \textit{et al.}, Physical Review C \textbf{86}, 054604 (2012).

\bibitem{elekes_16c} Z.\ Elekes, Z.\ Dombr\'adi, A.\ Krasznahorkay, H.\ Baba, M.\ Csatl\'os, L.\ Csige, N.\ Fukuda, Z.\ F\"ul\"op, Z.\ G\'acsi, J.\ Guly\'as, \textit{et al.}, Physics Letters B \textbf{586}, 34 (2004).

\bibitem{elekes_20c} Z.\ Elekes, Z.\ Dombr\'adi, T.\ Aiba, N.\ Aoi, H.\ Baba, D.\ Bemmerer, B.\ A.\ Brown, T.\ Furumoto, Z.\ F\"ul\"op, N.\ Iwasa, \textit{et al.}, Physical Review C \textbf{79}, 011302 (2009).

\bibitem{msu40si} C.\ M.\ Campbell, N.\ Aoi, D.\ Bazin, M.\ D.\ Bowen, B.\ A.\ Brown, J.\ M.\ Cook, D.-C.\ Dinca, A.\ Gade, T.\ Glasmacher, M.\ Horoi, \textit{et al.}, Physical Review Letters \textbf{97}, 112501 (2006).

\bibitem{stanoiu3} M.\ Stanoiu, D.\ Sohler, O.\ Sorlin, F.\ Azaiez, Z.\ Dombr\'adi, B.\ A.\ Brown, M.\ Belleguic, C.\ Borcea, C.\ Bourgeois, Z.\ Dlouhy, \textit{et al.}, Physical Review C \textbf{78}, 034315 (2008).

\bibitem{bowman} J.\ D.\ Bowman, A.\ M.\ Poskanzer, R.\ G.\ Korteling, and G.\ W.\ Butler, Physical Review C \textbf{9}, 836 (1974).

\bibitem{ozawa0} A.\ Ozawa, O.\ Bochkarev, L.\ Chulkov, D.\ Cortina, H.\ Geissel, M.\ Hellstr\"om, M.\ Ivanov, R.\ Janik, K.\ Kimura, T.\ Kobayashi, \textit{et al.}, Nuclear Physics A \textbf{691}, 599 (2001).

\bibitem{bazin} D.\ Bazin, B.\ A.\ Brown, J.\ Brown, M.\ Fauerbach, M.\ Hellstr\"om, S.\ E.\ Hirzebruch, J.\ H.\ Kelley, R.\ A.\ Kryger, D.\ J.\ Morrissey, R.\ Pfaff, \textit{et al.}, Physical Review Letters \textbf{74}, 3569 (1995).

\bibitem{maddalena} V.\ Maddalena, T.\ Aumann, D.\ Bazin, B.\ A.\ Brown, J.\ A.\ Caggiano, B.\ Davids, T.\ Glasmacher, P.\ G.\ Hansen, R.\ W.\ Ibbotson, A.\ Navin, \textit{et al.}, Physical Review C \textbf{63}, 024613 (2001).

\bibitem{baumann} T.\ Baumann, M.\ Borge, H.\ Geissel, H.\ Lenske, K.\ Markenroth, W.\ Schwab, M.\ Smedberg, T.\ Aumann, L.\ Axelsson, U.\ Bergmann, \textit{et al.}, Physics Letters B \textbf{439}, 256 (1998).

\bibitem{kanungo} R.\ Kanungo, Z.\ Elekes, H.\ Baba, Z.\ Dombr\"adi, Z.\ F\"ul\"op, J.\ Gibelin, A.\ Horv\'ath, Y.\ Ichikawa, E.\ Ideguchi, N.\ Iwasa, \textit{et al.}, Nuclear Physics A \textbf{757}, 315 (2005).

\bibitem{chiba} M.\ Chiba, R.\ Kanungo, B.\ Abu-Ibrahim, S.\ Adhikari, D.\ Fang, N.\ Iwasa, K.\ Kimura, K.\ Maeda, S.\ Nishimura, T.\ Ohnishi, \textit{et al.}, Nuclear Physics A \textbf{741}, 29 (2004).

\bibitem{marques} F.\ Marqu\'es, E.\ Liegard, N.\ Orr, J.\ Ang\'elique, L.\ Axelsson, G.\ Bizard, W.\ Catford, N.\ Clarke, G.\ Costa, M.\ Freer, \textit{et al.}, Physics Letters B \textbf{381}, 407 (1996).

\bibitem{yamaguchi} T.\ Yamaguchi, K.\ Tanaka, T.\ Suzuki, A.\ Ozawa, T.\ Ohtsubo, T.\ Aiba, N.\ Aoi, H.\ Baba, M.\ Fukuda, Y.\ Hashizume, \textit{et al.}, Nuclear Physics A \textbf{864}, 1 (2011).

\bibitem{elekes} Z.\ Elekes, Z.\ Dombr\'adi, R.\ Kanungo, H.\ Baba, Z.\ F\"ul\"op, J.\ Gibelin, A.\ Horv\'ath, E.\ Ideguchi, Y.\ Ichikawa, N.\ Iwasa, \textit{et al.}, Physics Letters B \textbf{614}, 174 (2005).

\bibitem{satou} Y.\ Satou, T.\ Nakamura, N.\ Fukuda, T.\ Sugimoto, Y.\ Kondo, N.\ Matsui, Y.\ Hashimoto, T.\ Nakabayashi, T.\ Okumura, M.\ Shinohara, \textit{et al.}, Physics Letters B \textbf{660}, 320 (2008).

\bibitem{thoennessen} M.\ Thoennessen, S.\ Mosby, N.\ Badger, T.\ Baumann, D.\ Bazin, M.\ Bennett, J.\ Brown, G.\ Christian, P.\ DeYoung, J.\ Finck, \textit{et al.}, Nuclear Physics A \textbf{912}, 1 (2013).

\bibitem{ozawa} A.\ Ozawa, Y.\ Hashizume, Y.\ Aoki, K.\ Tanaka, T.\ Aiba, N.\ Aoi, H.\ Baba, B.\ A.\ Brown, M.\ Fukuda, K.\ Inafuku, \textit{et al.}, Physical Review C \textbf{84}, 064315 (2011).

\bibitem{kobayashi} N.\ Kobayashi, T.\ Nakamura, J.\ A.\ Tostevin, Y.\ Kondo, N.\ Aoi, H.\ Baba, S.\ Deguchi, J.\ Gibelin, M.\ Ishihara, Y.\ Kawada, \textit{et al.}, Physical Review C \textbf{86}, 054604 (2012).

\bibitem{rips} T.\ Kubo, M.\ Ishihara, N.\ Inabe, H.\ Kumagai, I.\ Tanihata, K.\ Yoshida, T.\ Nakamura, H.\ Okuno, S.\ Shimoura, and K.\ Asahi, Nuclear Instruments and Methods in Physics Research Section B: Beam Interactions with Materials and Atoms \textbf{70}, 309 (1992).

\bibitem{pid} H.\ Sakurai, S.\ M.\ Lukyanov, M.\ Notani, N.\ Aoi, D.\ Beaumel, N.\ Fukuda, M.\ Hirai, E.\ Ideguchi, N.\ Imai, M.\ Ishihara, \textit{et al.}, Physics Letters B \textbf{448}, 180 (1999).

\bibitem{htarget} H.\ Ryuto, M.\ Kunibu, T.\ Minemura, T.\ Motobayashi, K.\ Sagara, S.\ Shimoura, M.\ Tamaki, Y.\ Yanagisawa, and Y.\ Yano, Nuclear Instruments and Methods in Physics Research Section A: Accelerators, Spectrometers, Detectors and Associated Equipment \textbf{555}, 1 (2005).

\bibitem{dali} S.\ Takeuchi, T.\ Motobayashi, Y.\ Togano, M.\ Matsushita, N.\ Aoi, K.\ Demichi, H.\ Hasegawa, and H.\ Murakami, Nuclear Instruments and Methods in Physics Research Section A: Accelerators, Spectrometers, Detectors and Associated Equipment \textbf{763}, 596 (2014).

\bibitem{geant4} S.\ Agostinelli, J.\ Allison, K.\ Amako, J.\ Apostolakis, H.\ Araujo, P.\ Arce, M.\ Asai, D.\ Axen, S.\ Banerjee, G.\ Barrand, \textit{et al.}, Nuclear Instruments and Methods in Physics Research Section A: Accelerators, Spectrometers, Detectors and Associated Equipment \textbf{506}, 250 (2003).

\bibitem{warburton} E.\ K.\ Warburton and B.\ A.\ Brown, Physical Review C \textbf{46}, 923 (1992).

\bibitem{dieperink} A.\ E.\ L.\ Dieperink and T.\ d.\ Forest, Physical Review C \textbf{10}, 543 (1974).

\end{thebibliography}

\end{document}